\documentclass{appolb}
\usepackage{epsfig}

\usepackage{graphicx}
\usepackage{subfigure}
\usepackage{bm} 
\begin{document}

\title{Kinetic description of mixtures of anisotropic fluids
\thanks{email addresses: wojciech.florkowski@ifj.edu.pl and oskar.madetko@ujk.edu.pl}
}
\author{Wojciech Florkowski
\address{The H. Niewodnicza\'nski Institute of Nuclear Physics, Polish Academy of Sciences, PL-31342 Krak\'ow, Poland, and \\
Institute of Physics, Jan Kochanowski University, PL-25406~Kielce, Poland}
\and
Oskar Madetko
\address{Institute of Physics, Jan Kochanowski University, PL-25406~Kielce, Poland}
}
\maketitle
\begin{abstract}
A simple system of coupled kinetic equations for quark and gluon anisotropic systems is solved numerically. The solutions are compared with the predictions of the anisotropic hydrodynamics describing a mixture of anisotropic fluids. We find that the solutions of the kinetic equations can be well reproduced by anisotropic hydrodynamics if the initial distribution are oblate for both quarks and gluons. On the other hand, the solutions of the kinetic equations have a different qualitative behavior from those obtained in anisotropic hydrodynamics if the initial configurations are oblate-prolate or prolate-prolate. This suggests that an extension of the anisotropic hydrodynamics scheme for the mixture of anisotropic fluids is needed, where higher moments of the kinetic equations are used and present simplifications are avoided.
\end{abstract}

PACS numbers: 12.38.Mh, 24.10.Nz, 25.75.-q, 51.10.+y, 52.27.Ny
  
\section{Introduction}

Soft hadronic matter production in ultrarelativistic heavy-ion collisions is most successfully described within the framework of  relativistic viscous hydrodynamics \cite{Israel:1979wp,Muronga:2003ta,Baier:2006um,
Romatschke:2007mq,Dusling:2007gi,Luzum:2008cw,
Song:2008hj,Denicol:2010tr,Schenke:2011tv,
Shen:2011eg,Bozek:2011wa,Niemi:2011ix,
Bozek:2012qs}. Nevertheless, the presence of large gradients at the very early stages of the collisions leads to conceptual problems connected with the application of viscous hydrodynamics. Even if the viscosity to entropy ratio is very small, large gradients imply substantial corrections to the equilibrium pressure \cite{Heller:2011ju,
Heller:2012je}. This, in turn, leads to high anisotropy of the produced system in the momentum space~\cite{Martinez:2009mf}. The energy-momentum tensor of viscous fluid becomes significantly different from the tensor describing the perfect fluid. Since the standard viscous hydrodynamics approach is based on a linearization around an isotropic background, large pressure anisotropies pose a problem, especially for the 2nd-order viscous hydrodynamics.

This type of problems have stimulated new theoretical developments which propose a reorganization of the viscous hydrodynamics expansion. The new approach, referred to as {\it anisotropic hydrodynamics}, incorporates the possibility of large momentum-space anisotropy at leading order  \cite{Florkowski:2010cf,Martinez:2010sc,
Ryblewski:2010bs,Martinez:2010sd,
Ryblewski:2011aq,Martinez:2012tu,
Ryblewski:2012rr}. The newest developments in anisotropic hydrodynamics include the calculation of the next-to-leading corrections to the spheroidal ansatz for the leading order distribution function \cite{Bazow:2013ifa} and generalization of the spheroidal ansatz to the ellipsoidal one in the leading order \cite{Tinti:2013vba}.
For a recent discussion of the concepts of anisotropic hydrodynamics see Ref.~\cite{Strickland:2014eua}.

Recently, it has been also demonstrated that for one-dimensional and boost-invariant simple (one-component) systems the results of anisotropic hydrodynamics agree very well with the predictions of kinetic theory \cite{Florkowski:2013lza,Florkowski:2013lya}. In this paper we continue this type of investigations and confront the results of the kinetic theory with the results of anisotropic hydrodynamics for mixtures of anisotropic fluids introduced in Refs.~\cite{Florkowski:2012as,Florkowski:2013uqa}.

The paper is organized as follows: In Sec.~\ref{sect:qgeqs} we introduce the kinetic equations. In Sec.~\ref{sect:baryon} we implement the baryon number conservation into our model. Sec.~\ref{sect:LM} discusses the Landau matching condition which couples the kinetic equations for quarks, antiquarks and gluons, and is used to determine the effective temperature of our system. In Sec.~\ref{sect:pressures} we present the formulas used to calculate the longitudinal and transverse pressures in the kinetic approach. In Sec.~\ref{sect:ahydro} we introduce the equations of anisotropic hydrodynamics.  Our numerical results are presented in Sec.~\ref{sect:res}. The paper is closed with summary and conclusions in Sec.~\ref{sect:concl}.

%%%%%%%%%%%%%%%%%%%%%%%%%%%%%%%%%%%%%%%%%%%
\section{Quark, antiquark and gluon kinetic equations}
\label{sect:qgeqs}
%%%%%%%%%%%%%%%%%%%%%%%%%%%%%%%%%%%%%%%%%%%%

In this paper we consider simple boost-invariant and one-dimensional kinetic equations of the form
\begin{equation}
\label{kineq}
\frac{d f(\tau,w,p_T)}{d \tau}=\frac{f^{\rm eq}(\tau,w,p_T)-f(\tau,w,p_T)}{\tau_{\rm eq}},
\end{equation}
where $f(\tau,w,p_T)$ may be the phase-space distribution function of quarks \mbox{($f=Q^+$)}, antiquarks ($f=Q^-$) or gluons ($f=G$), $\tau=\sqrt{t^2-z^2}$ is the longitudinal invariant time, \mbox{$w=t p_L  - z E $} describes the longitudinal momentum, $p_T$ is the magnitude of the transverse momentum of particles, and $\tau_{\rm eq}$ is the equilibration time. The formal solutions of Eqs.~(\ref{kineq}) have the form (see \cite{Florkowski:2013lza, Florkowski:2013lya} and also \cite{Baym:1984np,Baym:1985tna}),
\begin{equation}
\label{formsolf}
f(\tau,w,p_T) = D(\tau,\tau_0) f_0(w,p_T) + \int_{\tau_0}^{\tau} \frac{d \tau'}{\tau_{\rm eq}}\  D(\tau,\tau') f^{\rm eq}(\tau^\prime,w,p_T),
\end{equation}
where we have introduced the damping function
\begin{equation}
D(\tau_2,\tau_1)=\mathrm{exp}\Bigg[
-\int_{\tau_1}^{\tau_2}
\frac{d\tau''}{\tau_{\rm eq}(\tau'')}\Bigg].
\label{Damp1}
\end{equation}
The function $D(\tau_2,\tau_1)$ satisfies the two differential relations
\begin{eqnarray}
\frac{\partial D(\tau_2,\tau_1)}{\partial \tau_2} 
= - \frac{D(\tau_2,\tau_1)}{\tau_{\rm eq}(\tau_2)}, \quad
\frac{\partial D(\tau_2,\tau_1)}{\partial \tau_1} 
=  \frac{D(\tau_2,\tau_1)}{\tau_{\rm eq}(\tau_1)},
\label{Damp2}
\end{eqnarray}
and converges to unity if the two arguments are the same, $D(\tau,\tau)=1$. These properties imply the identity
\begin{eqnarray}
1 = D(\tau,\tau_0) + \int_{\tau_0}^\tau
\frac{d\tau^\prime}{\tau_{\rm eq}(\tau^\prime)}
D(\tau,\tau^\prime).
\label{Damp3}
\end{eqnarray}
In our considerations we restrict ourselves to the case of the constant relaxation time and use the form
\begin{equation}
D(\tau_2,\tau_1)=\exp\left(-\frac{\tau_2-\tau_1}{\tau_{\rm eq}}\right).
\label{Damp4}
\end{equation}

For the initial distribution functions we take the Romatschke-Strickland form \cite{Romatschke:2003ms}
\begin{equation}
\label{RS}
f_0(w,p_T) = \frac{g_0}{4\pi^3} \exp\Bigg[
\frac{\lambda_0}{\Lambda_0} -\frac{\sqrt{(1+\xi_0)w^2+p_T^2\tau_0^2}}{\Lambda_0\tau_0}\Bigg].
\end{equation}
The parameter $\tau_0$ specifies the initial time, $\xi_0$ is the initial momentum anisotropy, and $\Lambda_0$ is the initial transverse-momentum scale. We use the same values of the parameters $\xi_0$ and $\Lambda_0$ for quarks and antiquarks, denoted below as $\xi_q$ and $\Lambda_q$. On the other hand, the values of the parameters $\xi_0$ and $\Lambda_0$ for gluons, denoted below as $\xi_g$ and $\Lambda_g$, might be different from those used for quarks. The parameter $\lambda_0$ is a generalization of the chemical potential. The values of $\lambda_0$ for quarks and antiquarks are opposite, and its value for gluons is zero, $\lambda_g=0$.

The equilibrium distributions functions are simply the Boltzmann distributions rewritten in the boost-invariant way, which gives
\begin{equation}
f_{\rm eq}(\tau,w,p_T) =  \frac{g_0}{4\pi^3} \exp\Bigg[\frac{\mu (\tau)}{T(\tau} -\frac{\sqrt{w^2+p_T^2\tau^2}}{T(\tau) \tau}\Bigg].
\label{eqdistr}
\end{equation} 
The parameter $T$ appearing in the equilibrium distribution functions (\ref{eqdistr}) is an effective temperature. We use the same value of $T$ for quarks, antiquarks and gluons. It is determined by the Landau matching. In this way, the quark, antiquark and gluon kinetic equations become a system of the three coupled equations. The parameter $\mu$ is the baryon chemical potential. It vanishes for gluons and has opposite values for quarks and antiquarks. Below we use the letter $\mu$ to denote the quark baryon chemical potential.

In Eqs.~(\ref{RS}) and (\ref{eqdistr}) the factor $g_0$ accounts for internal degrees of freedom other than spin. In the numerical calculations we use $g_0 = g_{q} = 6$ for quarks (color and isospin degrees of freedom) and $g_0 = g_{g} = 8$ for gluons (color degrees of freedom).

\section{Baryon number conservation}
\label{sect:baryon}

In this Section we analyze the baryon number conservation in our model. The baryon number density is calculated from the equation
\begin{eqnarray}
b(\tau) &=& \frac{1}{3} \int dP \, p\cdot u 
\left(Q^+(\tau,w,p_T)-Q^-(\tau,w,p_T)\right) 
\nonumber \\
&=&
\frac{1}{3 \tau} \int dP\ v \Bigg[D(\tau,\tau_0)
\Bigg(Q_0^+(w,p_T)-Q_0^-(w,p_T)\Bigg) 
\nonumber\\ 
&+& \int_{\tau_0}^{\tau} \frac{d \tau'}{\tau_{\rm eq}}\ D(\tau,\tau')
\Bigg(
Q_{\rm eq}^+(\tau^\prime,w,p_T)-
Q_{\rm eq}^-(\tau^\prime,w,p_T) \Bigg)\Bigg],
\label{b1}
\end{eqnarray}
where $u^\mu=(t,0,0,z)/\tau$ is the boost-invariant (Bjorken) flow characterizing the longitudinal expansion of our system and $dP$ is the element of the three-momentum space~\footnote{The boost-invariant variables $w$ and $v$ were introduced in Refs.~\cite{Bialas:1984wv,Bialas:1984ap}.}
\begin{eqnarray}
dP = \frac{d^3p}{E} = \frac{dw\,d^2p_T}{v}, \quad v= \tau\, p\cdot u = \sqrt{w^2 + p_T^2 \tau^2}.
\label{dP}
\end{eqnarray}
A straightforward calculation where the variable $w$ is rescaled by the factor $\sqrt{1+\xi_q}$ leads to the expression
\begin{eqnarray}
b(\tau) = \frac{4 g_q}{3\pi^2}\ \Bigg[
\,D(\tau,\tau_0) \sinh \left( \frac{\lambda_q}{\Lambda_q} \right) 
\frac{\tau_0}{\tau} 
 \frac{\Lambda_q^3}{\sqrt{1+\xi_q}} 
\nonumber \\
+\int_{\tau_0}^{\tau} \frac{d\tau'}{\tau_{\rm eq}}\ \,D(\tau,\tau') \sinh \left( \frac{\mu^\prime}{T^\prime} \right) \frac{\tau'}{\tau} \,
 T^{\prime \,3} \Bigg].
 \label{b2}
\end{eqnarray}
In our one-dimensional boost-invariant model the baryon number conservation law, $\partial_\mu(b u^\mu)=0$, leads to the equation
\begin{eqnarray}
\frac{db(\tau)}{d\tau}+ \frac{b(\tau)}{\tau} = 0,
\label{barcon1}
\end{eqnarray}
which has the scaling solution
\begin{eqnarray}
b(\tau) = \frac{b_0 \tau_0}{\tau}.
\label{barcon2}
\end{eqnarray}
The parameter $b_0$ is the initial baryon number density given by the expression
\begin{eqnarray}
b_0 = \frac{4 g_q}{3 \pi^2} 
\sinh\left(\frac{\lambda_q}{\Lambda_q}\right) \frac{\Lambda_q^3}{\sqrt{1+\xi_q}}.
\end{eqnarray}
One may check, for details see Ref.~\cite{Florkowski:2012as}, that the baryon number is conserved if the baryon density $b(\tau)$ is equal to the baryon density of the equilibrium background $b_{\rm eq}(\tau)$, 
\begin{eqnarray}
b = b_{\rm eq} = \frac{4 g_q}{3 \pi^2} 
\sinh\left(\frac{\mu}{T}\right) T^3.
\label{beq}
\end{eqnarray}
Using Eqs.~(\ref{barcon2}) and (\ref{beq}) in Eq.~(\ref{b2}) we find that Eq.~(\ref{b2}) is automatically fulfilled due to the identity (\ref{Damp3}). This suggests that we should use the initial baryon density $b_0$ as a free parameter of our model and the ratio $\mu/T$ should be always calculated directly from Eq.~(\ref{beq}), namely
\begin{eqnarray}
\frac{\mu}{T}=\sinh^{-1}\left(
\frac{3\pi^2 b_0 \tau_0}{4 g_q T^3 \tau} \right).
\label{mudT}
\end{eqnarray}
In this way, we guarantee that the baryon number is conserved in our model. In addition to (\ref{mudT}) it is convenient to use the formula
\begin{eqnarray}
\frac{\lambda_q}{\Lambda_q}=\sinh^{-1}\left(
\frac{3\pi^2 b_0 \sqrt{1+\xi_q}}{4 g_q \Lambda_q^3} \right).
\label{lL}
\end{eqnarray}
With the help of Eq.~(\ref{lL}) we may eliminate the parameter $\lambda_q$ from our considerations and treat $b_0$, $\Lambda_q$, and $\xi_q$ as initial parameters.

%%%%%%%%%%%%%%%%%%%%%%%%%%%%%%%%%%%%
\section{Landau matching}
\label{sect:LM}
%%%%%%%%%%%%%%%%%%%%%%%%%%%%%%%%%%%%

The phase-space distribution functions may be used to obtain further characteristics of our system. In particular, the energy density is obtained as the integral
\begin{equation}
\label{endens}
\varepsilon = \frac{1}{\tau^2} \int dP\  v^2 \left[Q^+(\tau,w,p_T)+Q^-(\tau,w,p_T)
+G(\tau,w,p_T)\right].
\end{equation}
In the case of local equilibrium, one integrates the equilibrium distributions (\ref{eqdistr}) and obtains
\begin{equation}
\label{endenseq}
\varepsilon_{\rm eq} = \frac{6}{\pi^2} \left(2 g_{q} \cosh\frac{\mu}{T} + g_{g}\right) T^4.
\end{equation}
The Landau matching condition requires that the energy of the system is equal to the equilibrium energy density. In this way we obtain
\begin{eqnarray}
&& \frac{6}{\pi^2} \left(2 g_{q}  \cosh\frac{\mu(\tau)}{T(\tau)} + g_{g}\right) T^4(\tau) 
\nonumber \\
&& =  \frac{1}{\tau^2} \int dP\ v^2 \Bigg[D(\tau,\tau_0)
\Bigg(Q_0^+(w,p_T)+Q_0^-(w,p_T)+G_0(w,p_T)\Bigg) \label{LM} \\ 
&&  + \int_{\tau_0}^{\tau} \frac{d \tau'}{\tau_{\rm eq}}\ D(\tau,\tau^\prime)\Bigg(
Q_{\rm eq}^+(\tau^\prime,w,p_T)+
Q_{\rm eq}^-(\tau^\prime,w,p_T)+
G_{\rm eq}(\tau^\prime,w,p_T) \Bigg)\Bigg].
\nonumber
\end{eqnarray}
Using Eqs.~(\ref{RS}) and (\ref{eqdistr}) in Eq.~(\ref{LM}) we find
\begin{eqnarray}
&& \left(2 g_{q} \cosh\frac{\mu(\tau)}{T(\tau)} + g_{g}\right) T^4(\tau) 
\label{LM1} \\ 
&&  =
\frac{D(\tau,\tau_0)}{2} \Bigg[
2 g_{q} \cosh\left(\frac{\lambda_q}{\Lambda_q}\right) \Lambda_{q}^4\ \mathcal{H}\Bigg(\frac{\tau_0}{\tau}\frac{1}{\sqrt{1+\xi_{q}}}\Bigg) 
+ g_{g} \Lambda_{g}^4\ \mathcal{H}\Bigg(\frac{\tau_0}{\tau}\frac{1}{\sqrt{1+\xi_{g}}}\Bigg)\Bigg]
\nonumber\\ 
&&  + 
\int_{\tau_0}^{\tau} 
\frac{d\tau'}{2\tau_{\rm eq}}
D(\tau,\tau') 
\left(2 g_{q} \cosh\frac{\mu^\prime}{T^\prime} + g_{g}\right)\ 
T^{\prime \, 4} \,
\mathcal{H}\Bigg(\frac{\tau'}{\tau}\Bigg),
\nonumber
\end{eqnarray}
where we have introduced the function $\mathcal{H}$ defined by the integral~\footnote{The two integrations on the right-hand side of (\ref{LM}) may be done by changing to the polar coordinates. In the initial distributions we use the substitutions: $r \cos\phi = w \sqrt{1+\xi_0}/(\Lambda_0 \tau_0)$ and $r \sin\phi = p_T/\Lambda_0$. Similarly, in the thermal (integral) part we use: $r \cos\phi = w/(T^\prime \tau^\prime)$ and $r\sin\phi = p_T/T^\prime$.}
\begin{eqnarray}
 \mathcal{H}(y) = y\int_0^{\pi} d\phi\ \sin\phi\ \sqrt{y^2 \cos^2\phi + \sin^2\phi} =
 y \left( y + \frac{\cosh^{-1}y}{\sqrt{y^2-1}}\right).
\end{eqnarray}
By $T^\prime$ and $\mu^\prime$ we denote the values $T(\tau^\prime)$ and $\mu(\tau^\prime)$. We stress that the ratios $\mu/T$ and $\mu^\prime/T^\prime$ appearing in Eq.~(\ref{LM1}) should be expressed by the formula (\ref{mudT}). In this situation it is useful to introduce the notation
\begin{eqnarray}
\hspace{-1cm} h(b_0,T,\tau) &=& 2 g_{q} \cosh\left(
\frac{\mu}{T}\right) + g_{g}= 2 g_q \left[\left(
\frac{3\pi^2 b_0 \tau_0}{4 g_q T^3 \tau} \right)^2+1\right]^{1/2} +g_g
\label{h}
\end{eqnarray}
and
\begin{eqnarray}
{\bar g}_q = 2 g_q \cosh\left(\frac{\lambda_q}{\Lambda_q}\right) = 2 g_q \left[(1+\xi_q)
\left(
\frac{3\pi^2 b_0 }{4 g_q \Lambda_q^3} \right)^2
 +1
\right]^{1/2}.
\label{bargq}
\end{eqnarray}
Here we used the property of the hyperbolic functions $\cosh(\sinh^{-1}(x))=(x^2+1)^{1/2}$. For vanishing baryon number density, $b_0=0$, the function $h$ reduces to the total number of the internal degrees of freedom, $h(0,T,\tau)=h_0=2g_q+g_g$. Similarly, for $b_0=0$ the parameter ${\bar g}_q$ reduces to total number of the quark and antiquark degrees of freedom,  ${\bar g}_q=2g_q$.

The definitions (\ref{h}) and (\ref{bargq}) allow us to rewrite the Landau matching condition in a more compact form as
\begin{eqnarray}
&& h\left(b_0,T(\tau),\tau\right) T^4(\tau) 
\label{LM2} \\ 
&&  =
\frac{D(\tau,\tau_0)}{2} \Bigg[
{\bar g}_{q}  \Lambda_{q}^4\ \mathcal{H}\Bigg(\frac{\tau_0}{\tau}\frac{1}{\sqrt{1+\xi_{q}}}\Bigg) 
+ g_{g} \Lambda_{g}^4\ \mathcal{H}\Bigg(\frac{\tau_0}{\tau}\frac{1}{\sqrt{1+\xi_{g}}}\Bigg)\Bigg]
\nonumber\\ 
&&  \hspace{1cm} + 
\int_{\tau_0}^{\tau} 
\frac{d\tau'}{2\tau_{\rm eq}}
D(\tau,\tau') 
h(b_0,T^\prime,\tau^\prime) \ 
T^{\prime \, 4} \,
\mathcal{H}\Bigg(\frac{\tau'}{\tau}\Bigg),
\nonumber
\end{eqnarray}
We note that Eq.~(\ref{LM1}) is an integral equation for the function $T(\tau)$ that may be solved with the iterative method \cite{Banerjee:1989by}.

%%%%%%%%%%%%%%%%%%%%%%%%%%%%%%%%%%%%%%%%%%%%%%%%%
\section{Longitudinal and transverse pressures}
\label{sect:pressures}
%%%%%%%%%%%%%%%%%%%%%%%%%%%%%%%%%%%%%%%%%%%%%%%%%

Within the kinetic approach, the transverse and longitudinal pressures are obtained from the expressions
\begin{eqnarray}
P_T(\tau) &=& \frac{1}{2}\int dP\ p_T^2\
\Bigg[D(\tau,\tau_0) \left(Q_0^+(w,p_T)+Q_0^-(w,p_T)+G_0(w,p_T)\right)
\nonumber \\
&& \hspace{-1.5cm} 
+\int\frac{d\tau'}{\tau_{\rm eq}} D(\tau,\tau') \left(
Q_{\rm eq}^+(\tau^\prime,w,p_T)+
Q_{\rm eq}^-(\tau^\prime,w,p_T)+
G_{\rm eq}(\tau^\prime,w,p_T) \right)
\Bigg]
\label{PT}
\end{eqnarray}
and
\begin{eqnarray}
P_L(\tau) &=& \frac{1}{\tau^2}\int dP\ w^2\
\Bigg[D(\tau,\tau_0) \left(Q_0^+(w,p_T)+Q_0^-(w,p_T)+G_0(w,p_T)\right)
\nonumber \\
&& \hspace{-1.5cm} 
+\int\frac{d\tau'}{\tau_{\rm eq}} D(\tau,\tau') \left(
Q_{\rm eq}^+(\tau^\prime,w,p_T)+
Q_{\rm eq}^-(\tau^\prime,w,p_T)+
G_{\rm eq}(\tau^\prime,w,p_T) \right)
\Bigg].
\label{PL}
\end{eqnarray}
Equations (\ref{PT}) and (\ref{PL}) may be rewritten as
\begin{eqnarray}
P_T &=& \frac{3}{\pi^2}\Bigg[
\frac{D(\tau,\tau_0)}{2}\  
\left({\bar g}_q \Lambda_q^4 \mathcal{H}_T\Bigg(\frac{\tau_0}{\tau}\frac{1}{\sqrt{1+\xi_q}}\Bigg)
+ g_g \Lambda_g^4 \mathcal{H}_T\Bigg(\frac{\tau_0}{\tau}\frac{1}{\sqrt{1+\xi_g}}\Bigg)
\right)
\nonumber \\
&& \hspace{2cm} +  \int_{\tau_0}^{\tau}\frac{d\tau'}{2\tau_{\rm eq}}D(\tau,\tau') h(b_0,T^\prime,\tau^\prime) T'^4\ \mathcal{H}_T\Bigg(\frac{\tau'}{\tau}\Bigg)\Bigg]
\end{eqnarray}
and
\begin{eqnarray}
P_L &=& \frac{6}{\pi^2}\Bigg[
\frac{D(\tau,\tau_0)}{2}\  
\left({\bar g}_q \Lambda_q^4 \mathcal{H}_L\Bigg(\frac{\tau_0}{\tau}\frac{1}{\sqrt{1+\xi_q}}\Bigg)
+ g_g \Lambda_g^4 \mathcal{H}_L\Bigg(\frac{\tau_0}{\tau}\frac{1}{\sqrt{1+\xi_g}}\Bigg)
\right)
\nonumber \\
&& \hspace{2cm} +  \int_{\tau_0}^{\tau}\frac{d\tau'}{2\tau^{\rm eq}}D(\tau,\tau') h(b_0,T^\prime,\tau^\prime) T'^4\ \mathcal{H}_L\Bigg(\frac{\tau'}{\tau}\Bigg)\Bigg],
\end{eqnarray}
where we have introduced the two auxiliary functions 
\begin{eqnarray}
\hspace{-0.75cm}  \mathcal{H}_T(y) &=& y\int_0^{\pi} \frac{d\phi\,\sin^3\phi}{\sqrt{y^2\cos^2\phi+\sin^2\phi}}
= \frac{y}{1-y^2} \left(
y  + \frac{1- 2 y^2}{\sqrt{y^2-1}} \cosh^{-1}y \right)
\label{HT}
\end{eqnarray}
and
\begin{eqnarray}
\hspace{-0.75cm}  \mathcal{H}_L(y) &=& y^3\int_0^{\pi} \frac{d\phi\,\cos^2\phi\sin\phi}{\sqrt{y^2\cos^2\phi+\sin^2\phi}}
= \frac{y^3}{y^2-1} \left(y-
\frac{\cosh^{-1}y}{\sqrt{y^2-1}} \right).
\label{HL}
\end{eqnarray}
These functions have been originally introduced in Ref.~\cite{Florkowski:2013lza,Florkowski:2013lya}. The relation $\varepsilon=2P_T+P_L$ valid for massless particles considered in this paper as well as the explicit definitions of the functions ${\cal H}$ lead to the condition ${\cal H}={\cal H}_T+{\cal H}_L$.

There are simple relations connecting ${\cal H}$, ${\cal H}_L$, and ${\cal H}_T$ with the functions ${\cal R}$, ${\cal R}_L$, and ${\cal R}_T$ defined in Ref.~\cite{Martinez:2010sc}, namely
\begin{eqnarray}
{\cal H}\left(\frac{1}{\sqrt{1+\xi}}\right) &=& 2 \, \cal{R}(\xi),
\nonumber \\
{\cal H}_L\left(\frac{1}{\sqrt{1+\xi}}\right) &=& \frac{2}{3} {\cal R}_L(\xi),
\label{Rs} \\
{\cal H}_T\left(\frac{1}{\sqrt{1+\xi}}\right) &=& \frac{4}{3} {\cal R}_T(\xi). \nonumber
\end{eqnarray}
We shall use the functions ${\cal R}$ in the framework of anisotropic hydrodynamics introduced in the next Section.

%%%%%%%%%%%%%%%%%%%%%%%%%%%%%%%%%%%%%%%%%%%
\section{Anisotropic hydrodynamics for a mixture of fluids}
\label{sect:ahydro}
%%%%%%%%%%%%%%%%%%%%%%%%%%%%%%%%%%%%%%%%%%%%

The results of the kinetic calculations will be compared in the next Section to the results of anisotropic hydrodynamics for a quark-gluon mixture. Such hydrodynamic equations were derived in \cite{Florkowski:2012as} and for the vanishing baryon density they have the form
\begin{eqnarray}
&& \frac{3}{\Lambda} \frac{d\Lambda}{d\tau} -\frac{1}{2(1+\xi_q)} \frac{d\xi_q}{d\tau} + \frac{1}{\tau} = \frac{1}{\tau_{\rm eq}^{\rm AH}} \left[ \left(\frac{{\cal R}(\xi_q) + r {\cal R}(\xi_g) }{1+r} \right)^{3/4} \sqrt{1+\xi_q} -1 \right], \nonumber \\
&& \frac{3}{\Lambda} \frac{d\Lambda}{d\tau} -\frac{1}{2(1+\xi_g)}  \frac{d\xi_g}{d\tau}+ \frac{1}{\tau} = \frac{1}{\tau_{\rm eq}^{\rm AH}} \left[ \left(\frac{{\cal R}(\xi_q) + r {\cal R}(\xi_g) }{1+r} \right)^{3/4} \sqrt{1+\xi_g} -1 \right] ,
\nonumber \\
&& \frac{4}{\Lambda} \frac{d\Lambda}{d\tau}
\left({\cal R}(\xi_q) + r {\cal R}(\xi_g)\right)
+ {\cal R}^\prime(\xi_q)\frac{d\xi_q}{d\tau} 
+ r {\cal R}^\prime(\xi_g)  \frac{d\xi_g}{d\tau}
\nonumber \\
&& \hspace{3cm} =
\frac{2}{\tau} \left( (1+\xi_q)  {\cal R}^\prime(\xi_q) + r (1+\xi_g)  {\cal R}^\prime(\xi_g) \right).
\label{AH}
\end{eqnarray}
Here $\xi_q$ and $\xi_g$ are the anisotropy parameters for quarks and gluons, respectively, $\Lambda$ is the transverse-momentum scale, common for quarks and gluons, $r= g_g/(2g_q) = 2/3$ is the ratio of the internal degrees of freedom, and ${\cal R}$ is the function defined in the first line of  Eq.~(\ref{Rs}). The parameter $\tau_{\rm eq}^{\rm AH}$ is the relaxation time used in anisotropic hydrodynamics. In Ref.~\cite{Florkowski:2013lya}, where a simple (one-component) fluid was studied, it has been found that $\tau_{\rm eq}^{\rm AH}$ is related to the relaxation time used in the kinetic theory through the expression
\begin{eqnarray}
\tau_{\rm eq}^{\rm AH} = \frac{ \tau_{\rm eq} \,T}{2\Lambda}.
\label{taueqAH1}
\end{eqnarray}
In our case (see, for example, Eq. (44) in \cite{Florkowski:2012as}) Eq.~(\ref{taueqAH1}) takes the form
\begin{eqnarray}
\tau_{\rm eq}^{\rm AH} = \frac{\tau_{\rm eq}}{2}
\left( \frac{{\cal R}(\xi_q) + r {\cal R}(\xi_g)}{1+r} \right)^{1/4}.
\label{taueqAH2}
\end{eqnarray}

If the solutions of Eqs.~(\ref{AH}) are known, one may calculate the time dependence of the energy density and the two pressures from the following equations
\begin{eqnarray}
\varepsilon &=& \frac{6 \Lambda^4}{\pi^2}
\left[ 2 g_q {\cal R}(\xi_q) + g_g {\cal R}(\xi_g)
\right], \nonumber \\
P_T &=& \frac{2 \Lambda^4}{\pi^2}
\left[ 2 g_q {\cal R}_T(\xi_q) + g_g {\cal R}_T(\xi_g)
\right], \nonumber \\
P_L &=& \frac{2 \Lambda^4}{\pi^2}
\left[ 2 g_q {\cal R}_L(\xi_q) + g_g {\cal R}_L(\xi_g)
\right]. \nonumber \\
\label{hydroobs}
\end{eqnarray}
Since in the hydrodynamic calculations quarks and antiquarks are included together, there is an additional factor of 2 multiplying $g_q$ in (\ref{hydroobs}).

\begin{figure}[t]
\begin{center}
\subfigure{\includegraphics[angle=0,width=0.49\textwidth]{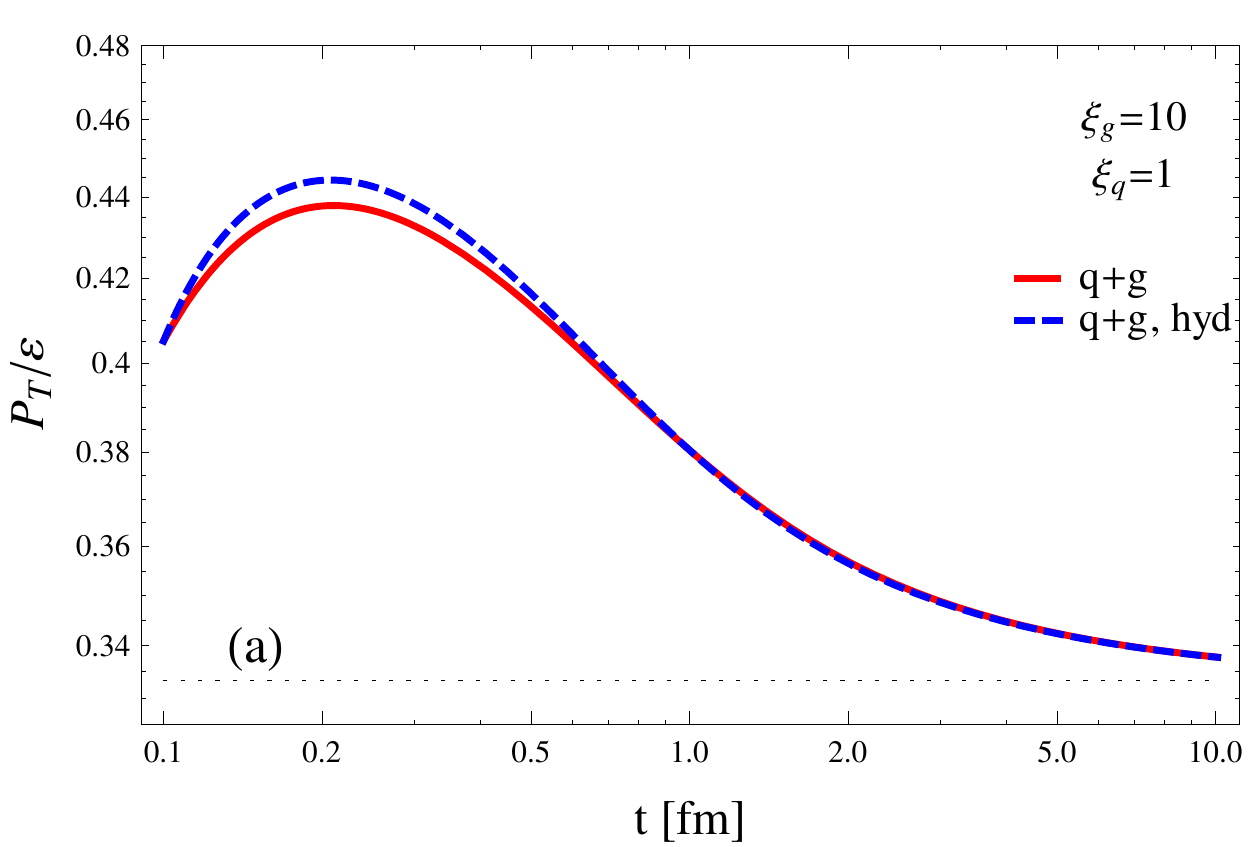}} 
\subfigure{\includegraphics[angle=0,width=0.49\textwidth]{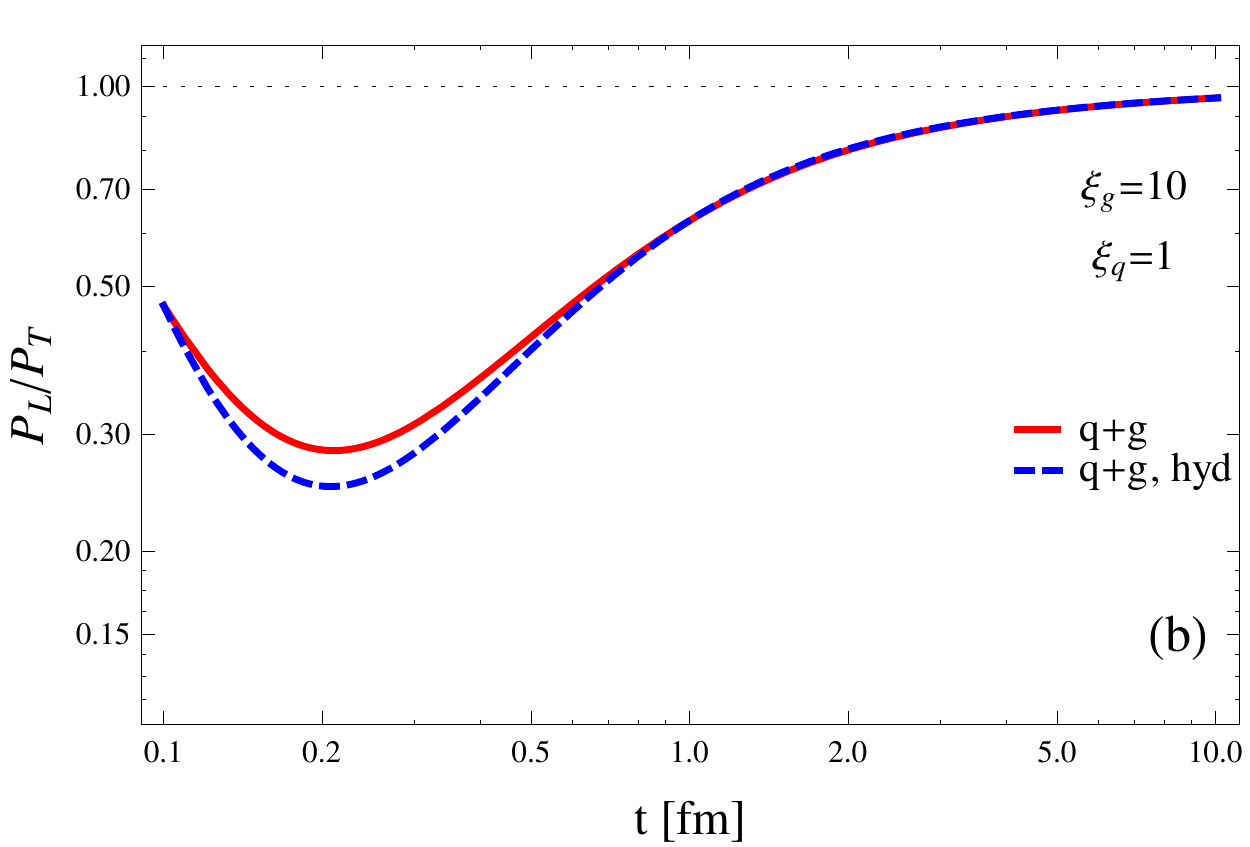}} \\
\subfigure{\includegraphics[angle=0,width=0.49\textwidth]{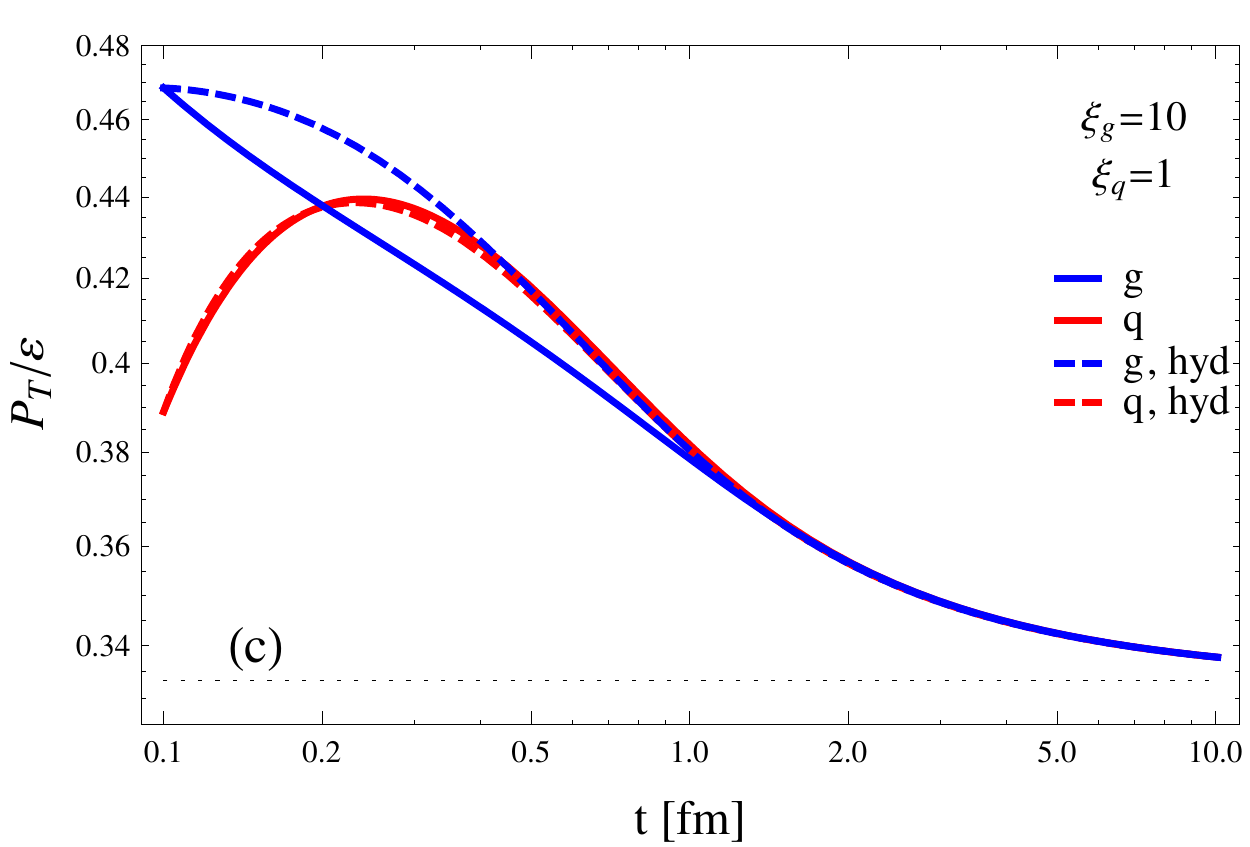}} 
\subfigure{\includegraphics[angle=0,width=0.49\textwidth]{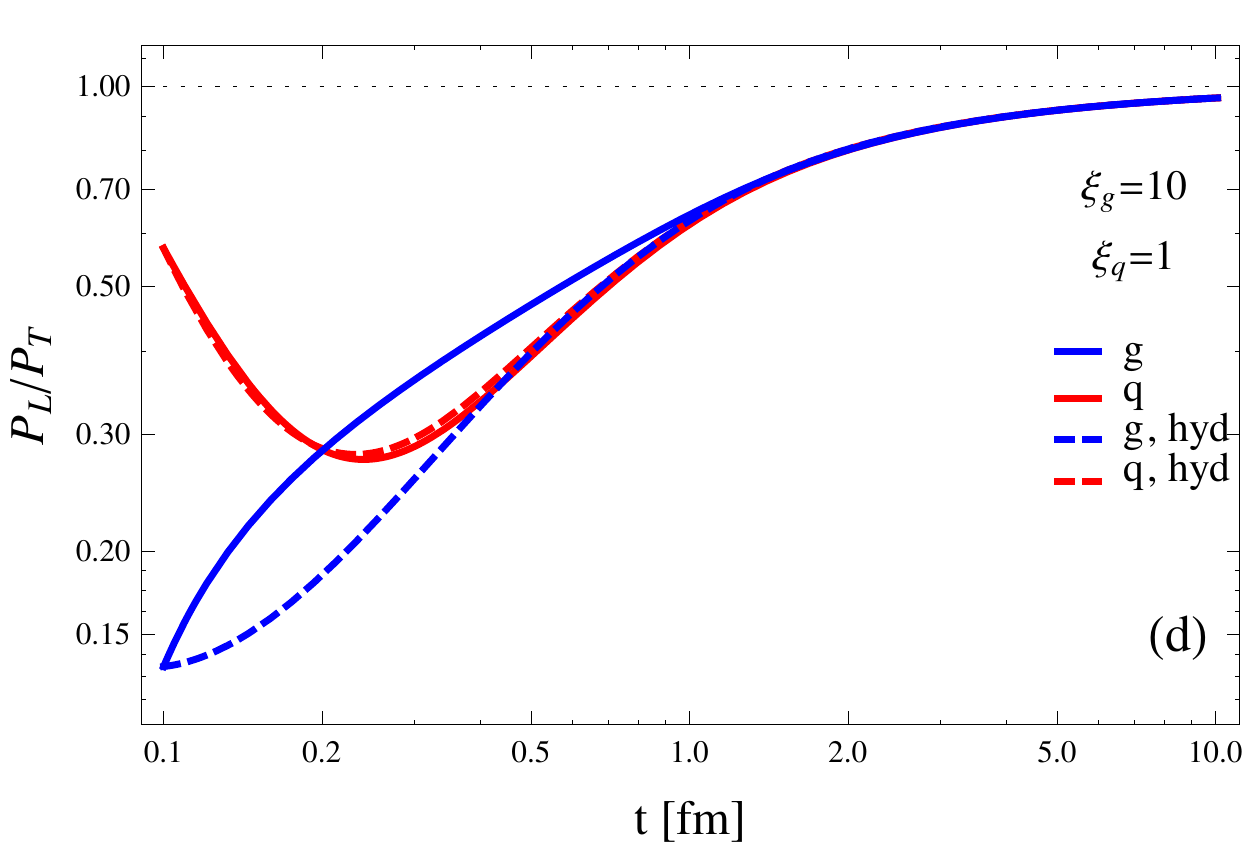}}
\end{center}
\caption{(Color online) Time evolution of the ratios $P_T/\varepsilon$ (left columns) and $P_L/P_T$ (right columns) for the system which is initially oblate-oblate, i.e., the two anisotropy parameters are positive, in this case $\xi_g=10$ and $\xi_q =1$. The two upper panels show the global characteristics of the system, while the two lower panels show the individual results for quarks and gluons, separately. The results of the kinetic calculation are represented by the solid lines, while the hydrodynamic results are represented by the dashed lines.
}
\label{fig:oo}
\end{figure}

\begin{figure}[t]
\begin{center}
\subfigure{\includegraphics[angle=0,width=0.49\textwidth]{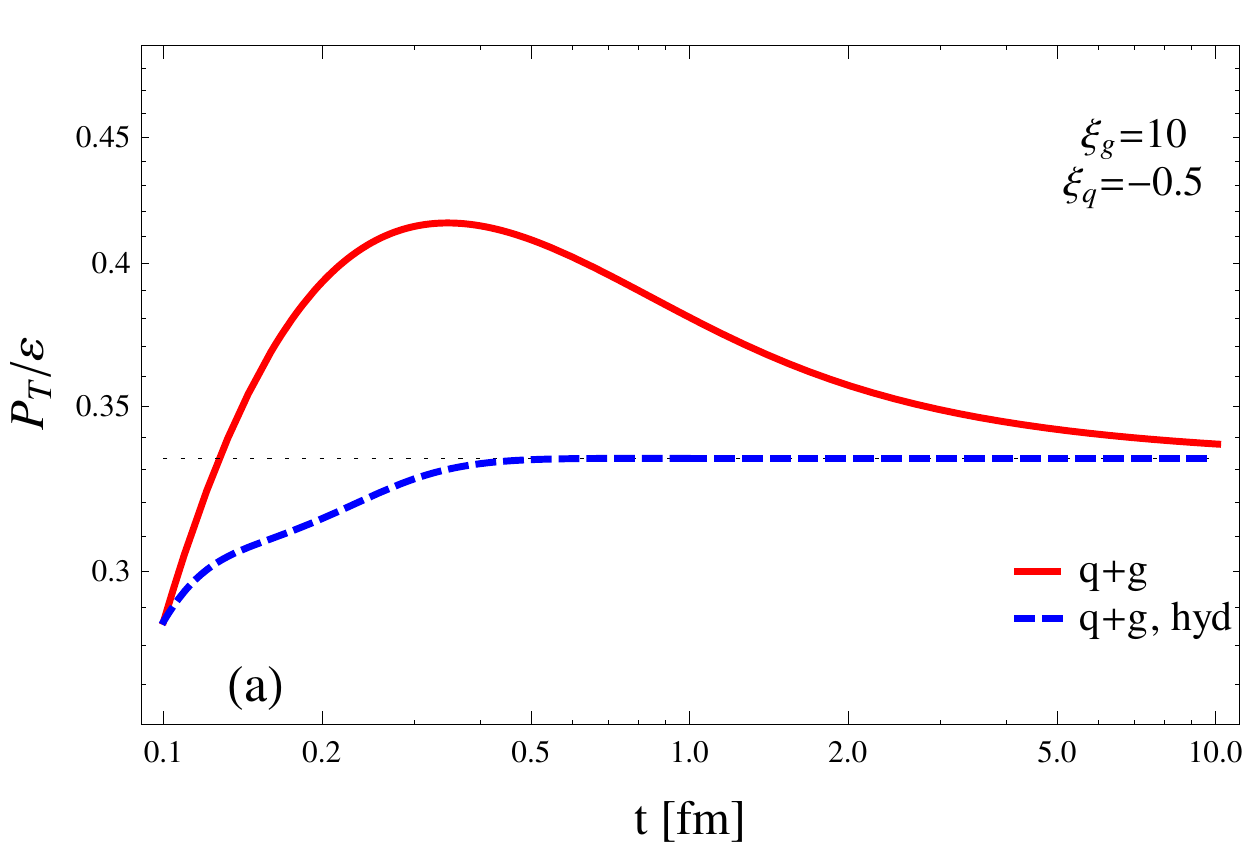}} 
\subfigure{\includegraphics[angle=0,width=0.49\textwidth]{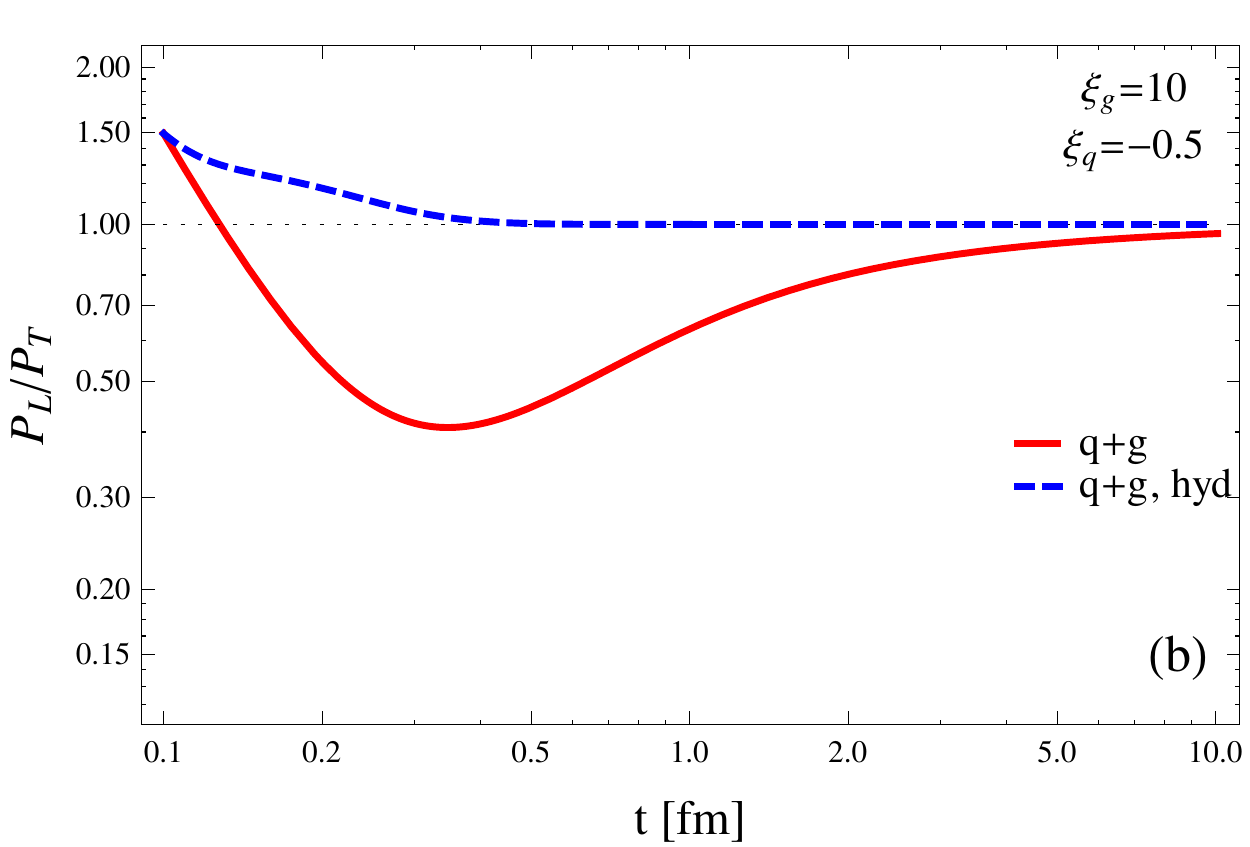}} \\
\subfigure{\includegraphics[angle=0,width=0.49\textwidth]{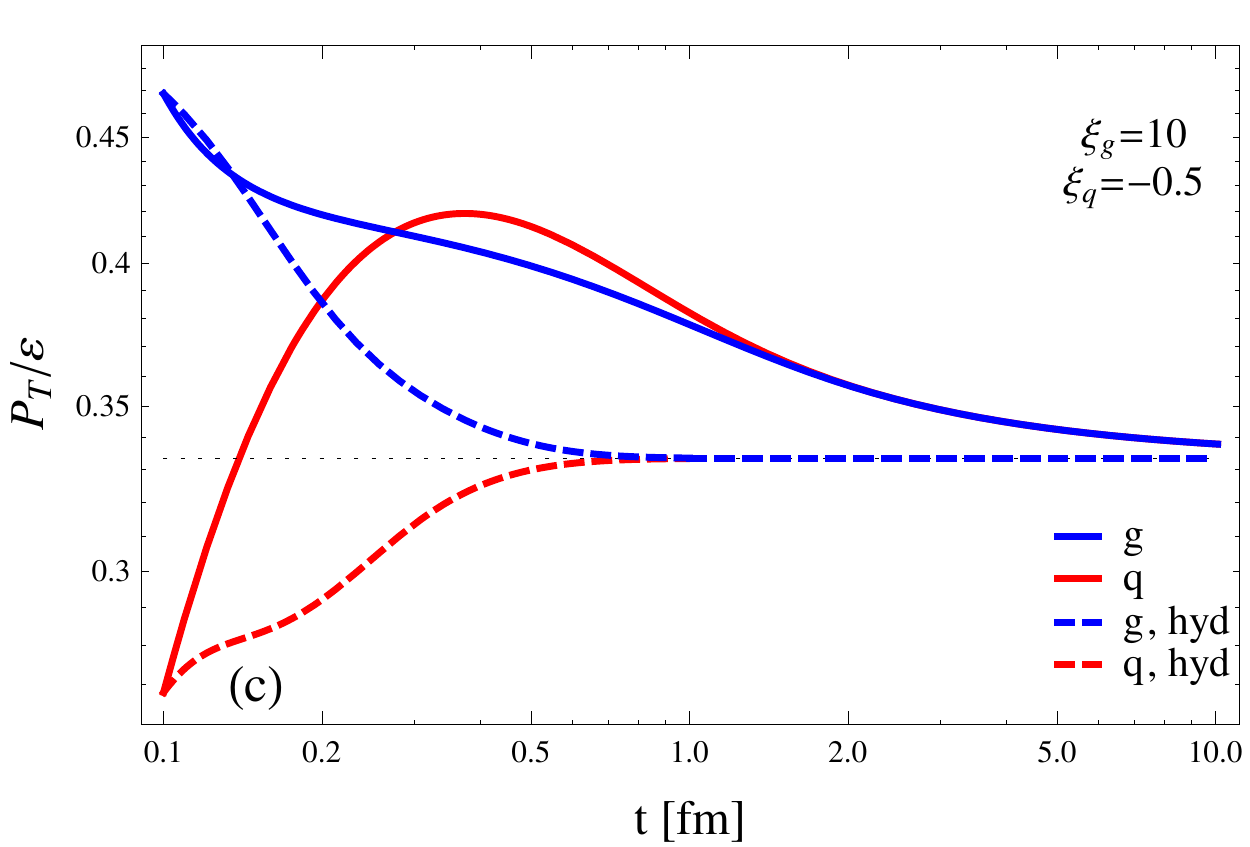}} 
\subfigure{\includegraphics[angle=0,width=0.49\textwidth]{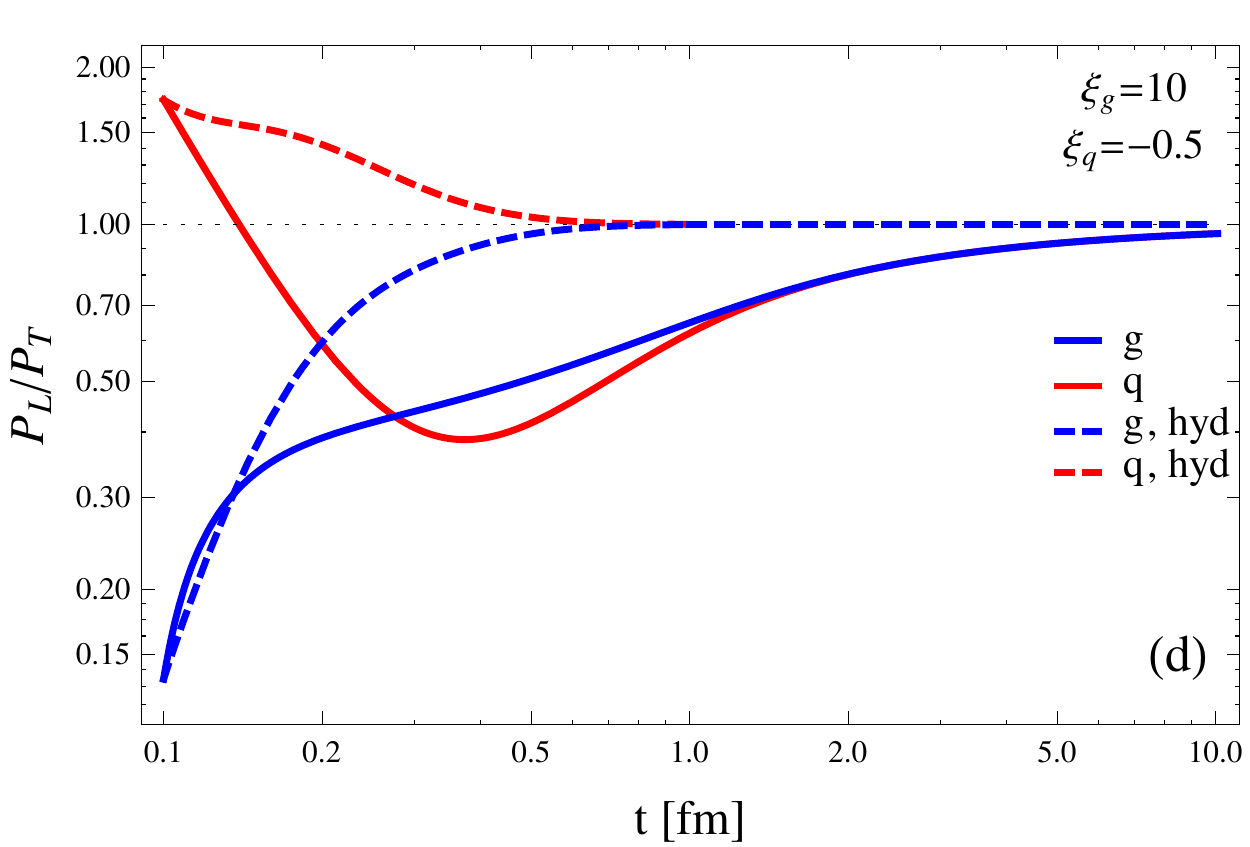}}
\end{center}
\caption{(Color online) The same as Fig.~\ref{fig:oo} but for the initial oblate-prolate configuration, $\xi_g =10$ and $\xi_q=-0.5$.
}
\label{fig:op}
\end{figure}

\begin{figure}[t]
\begin{center}
\subfigure{\includegraphics[angle=0,width=0.49\textwidth]{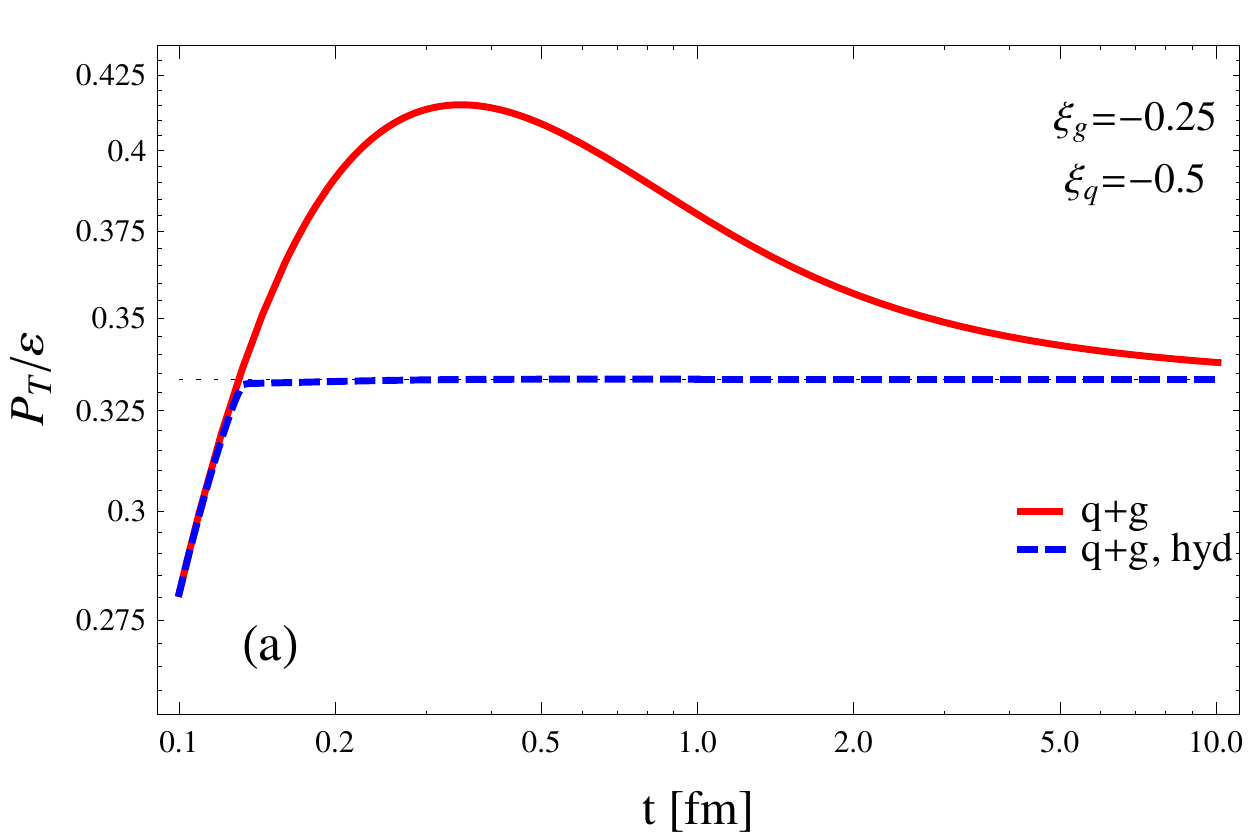}} 
\subfigure{\includegraphics[angle=0,width=0.49\textwidth]{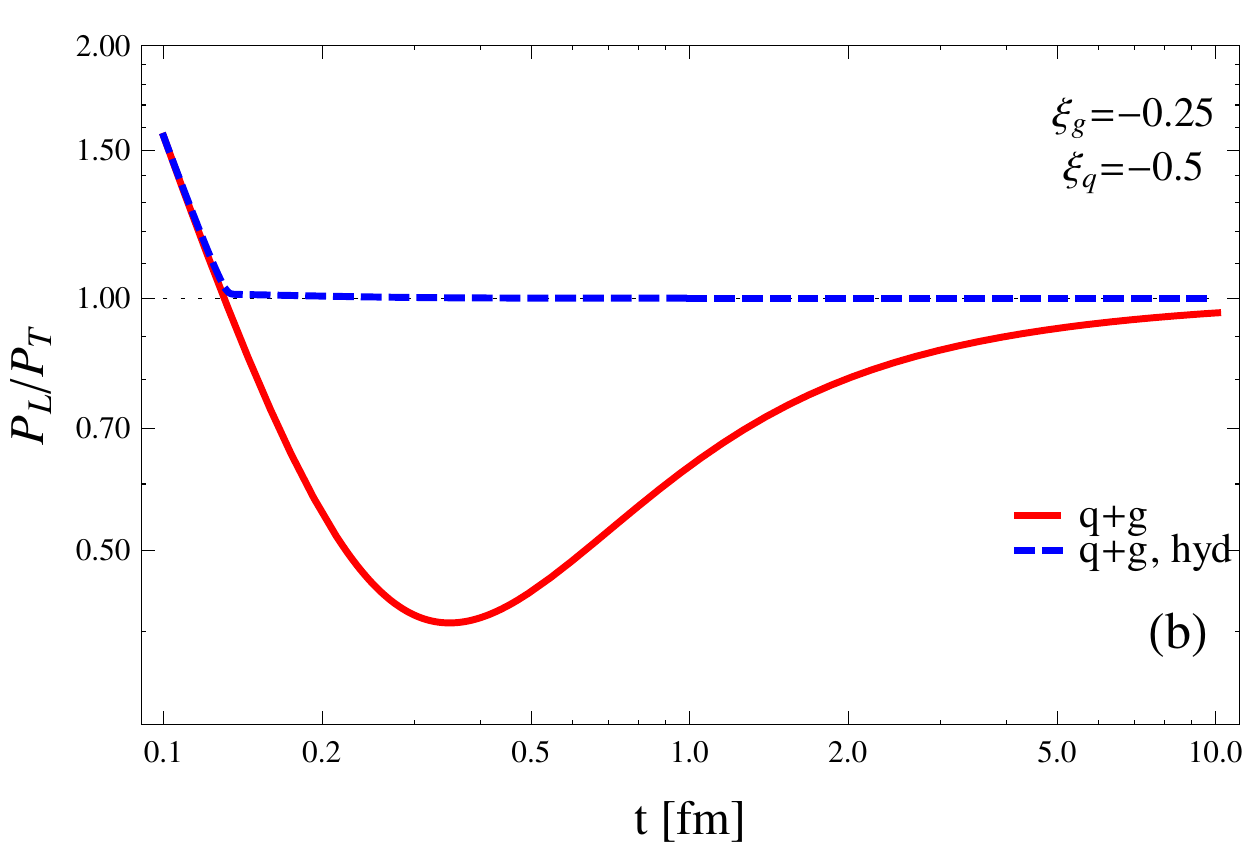}} \\
\subfigure{\includegraphics[angle=0,width=0.49\textwidth]{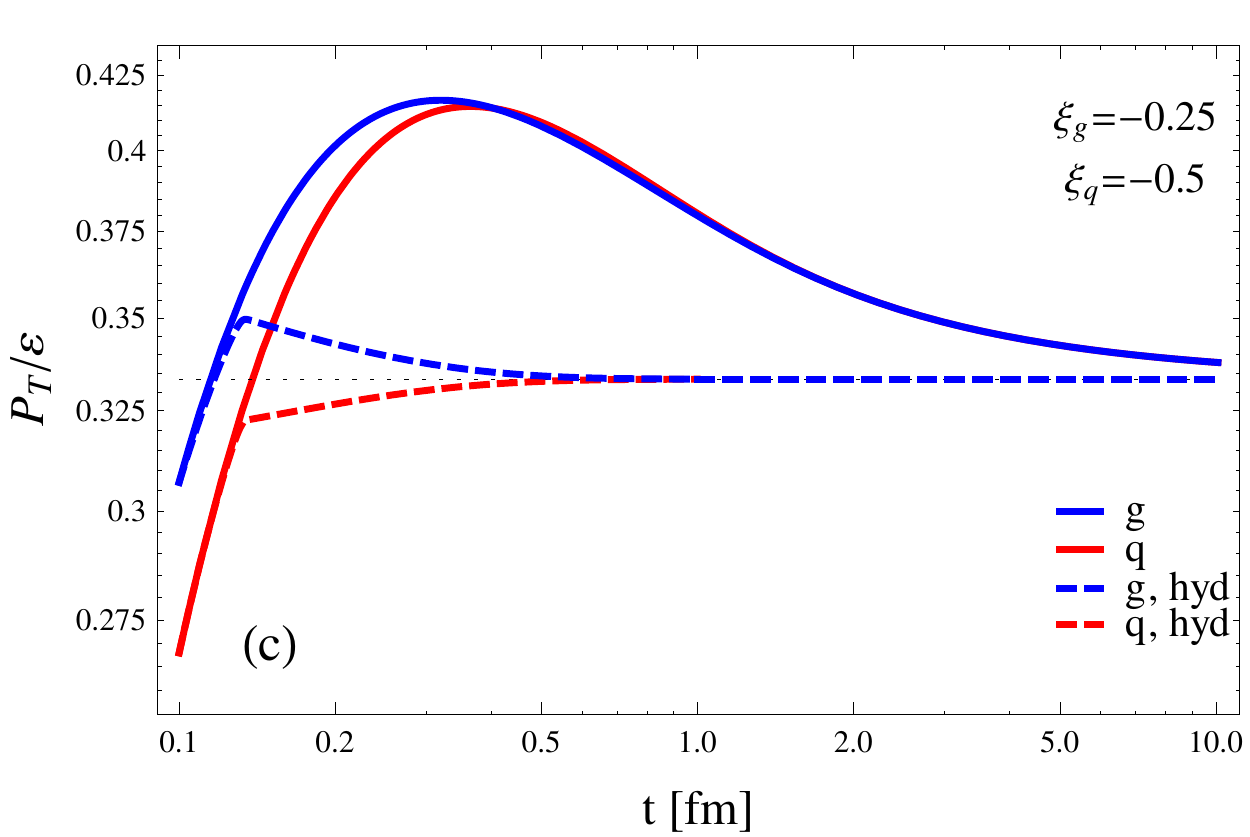}} 
\subfigure{\includegraphics[angle=0,width=0.49\textwidth]{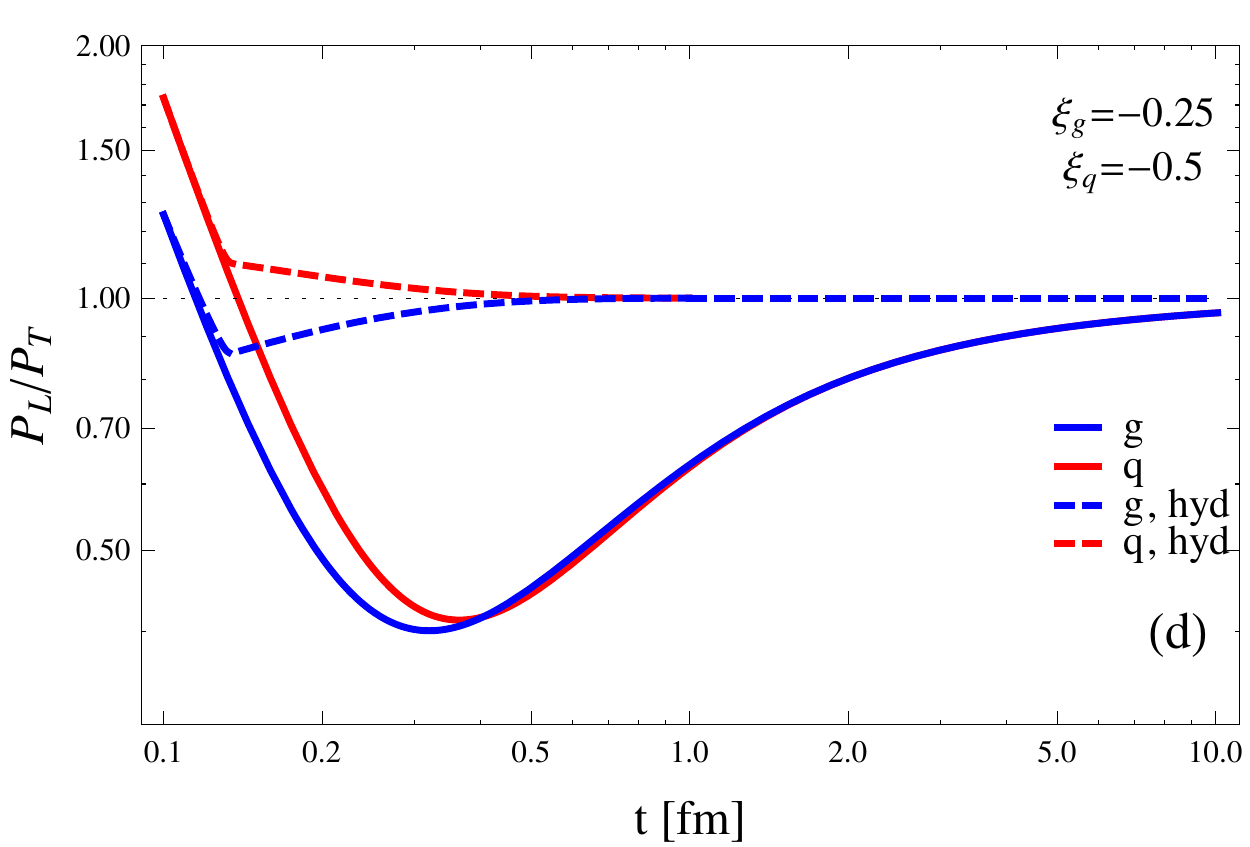}}
\end{center}
\caption{(Color online) The same as Fig.~\ref{fig:oo} but for the initial prolate-prolate configuration, $\xi_g =-0.25$ and $\xi_q=-0.5$.
}
\label{fig:pp}
\end{figure}

\begin{figure}[t]
\begin{center}
\subfigure{\includegraphics[angle=0,width=0.49\textwidth]{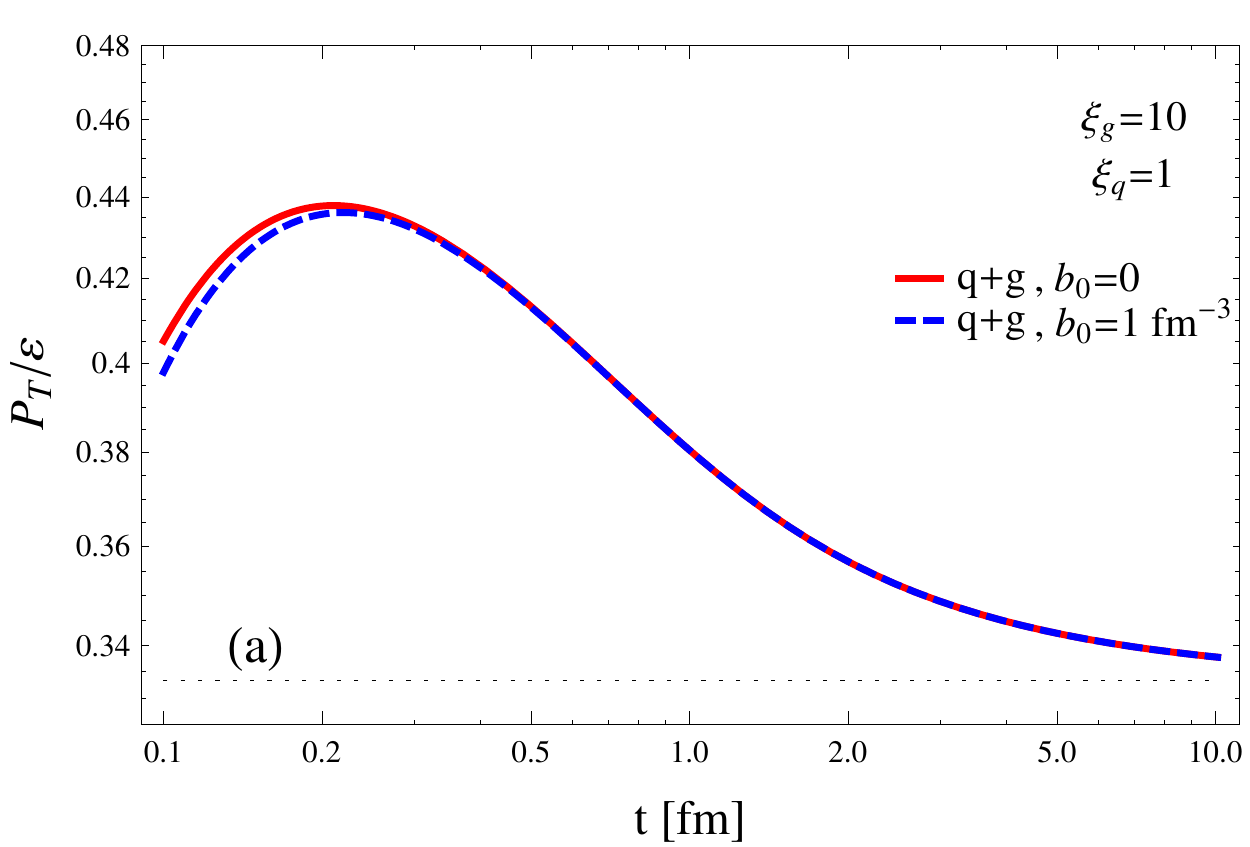}} 
\subfigure{\includegraphics[angle=0,width=0.49\textwidth]{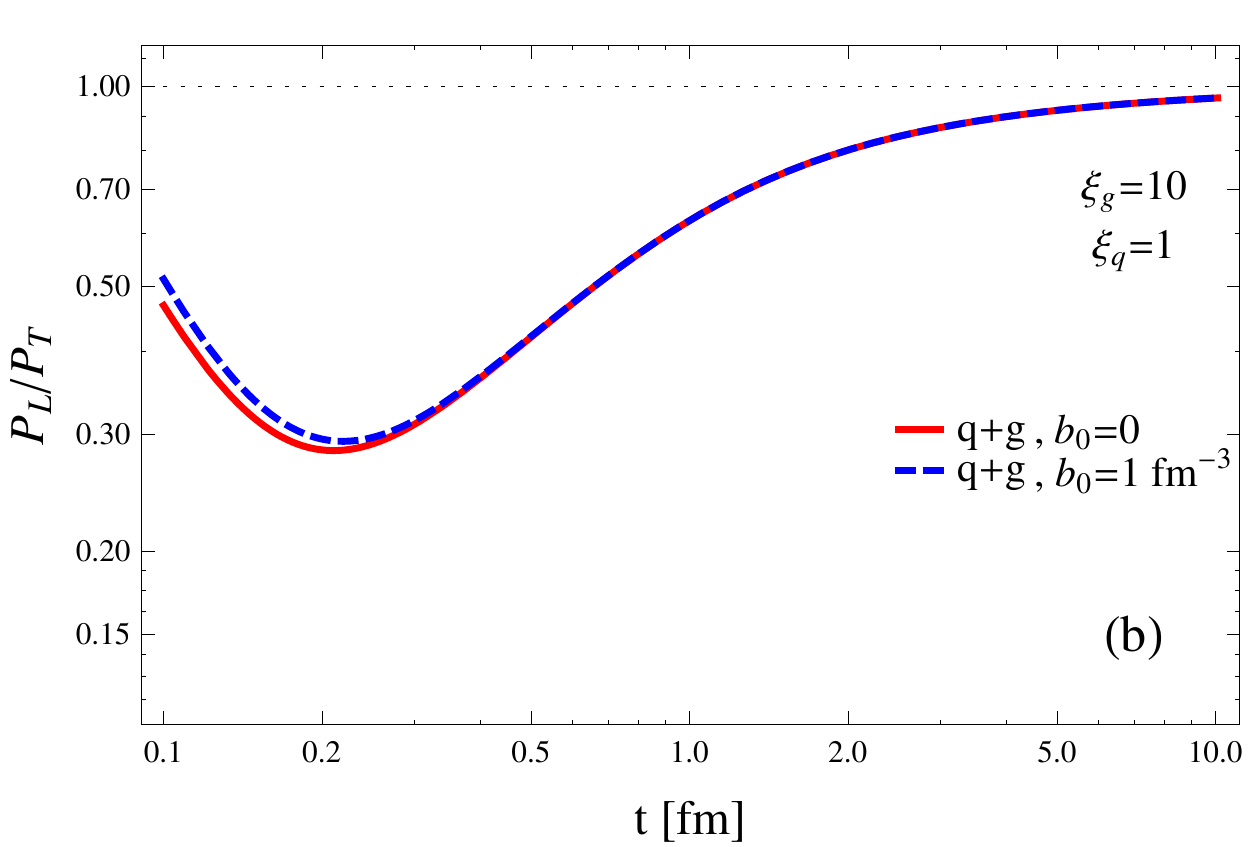}} \\
\subfigure{\includegraphics[angle=0,width=0.49\textwidth]{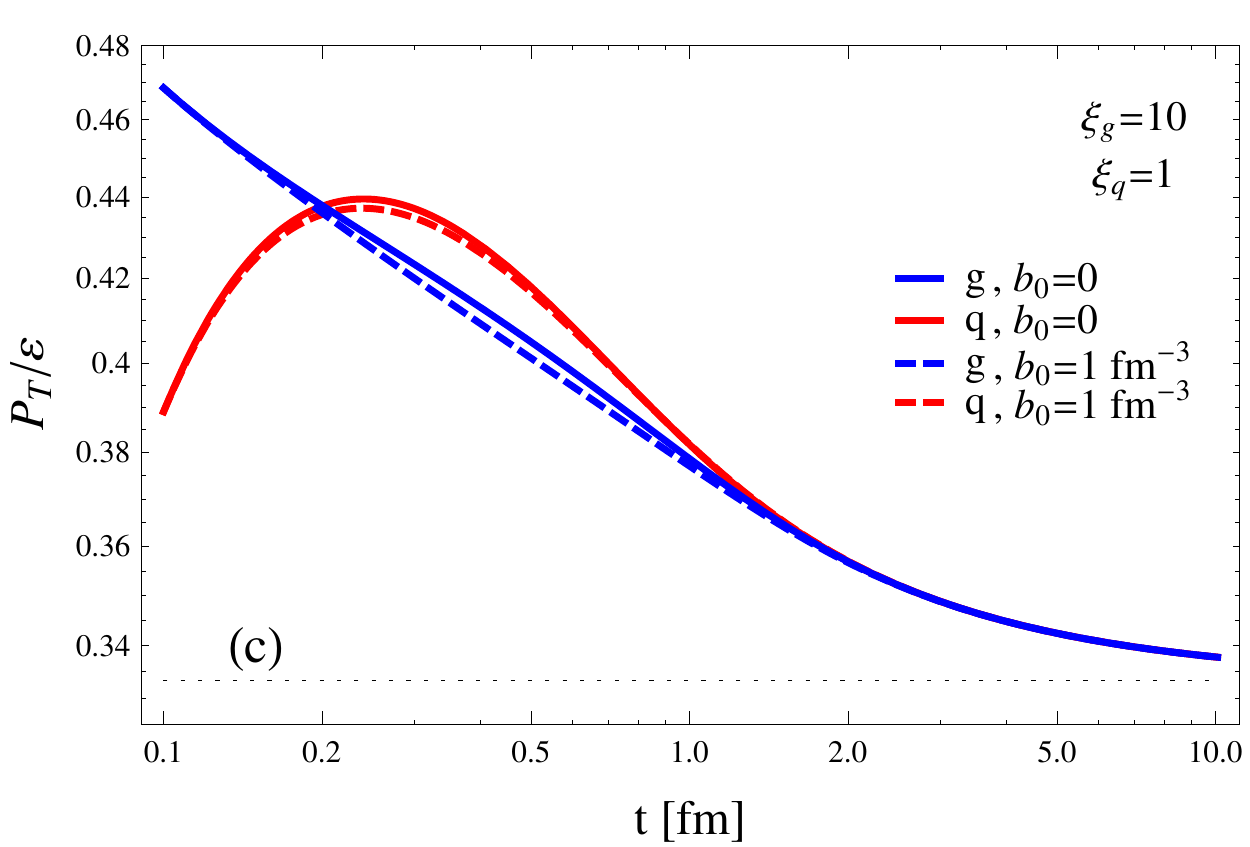}} 
\subfigure{\includegraphics[angle=0,width=0.49\textwidth]{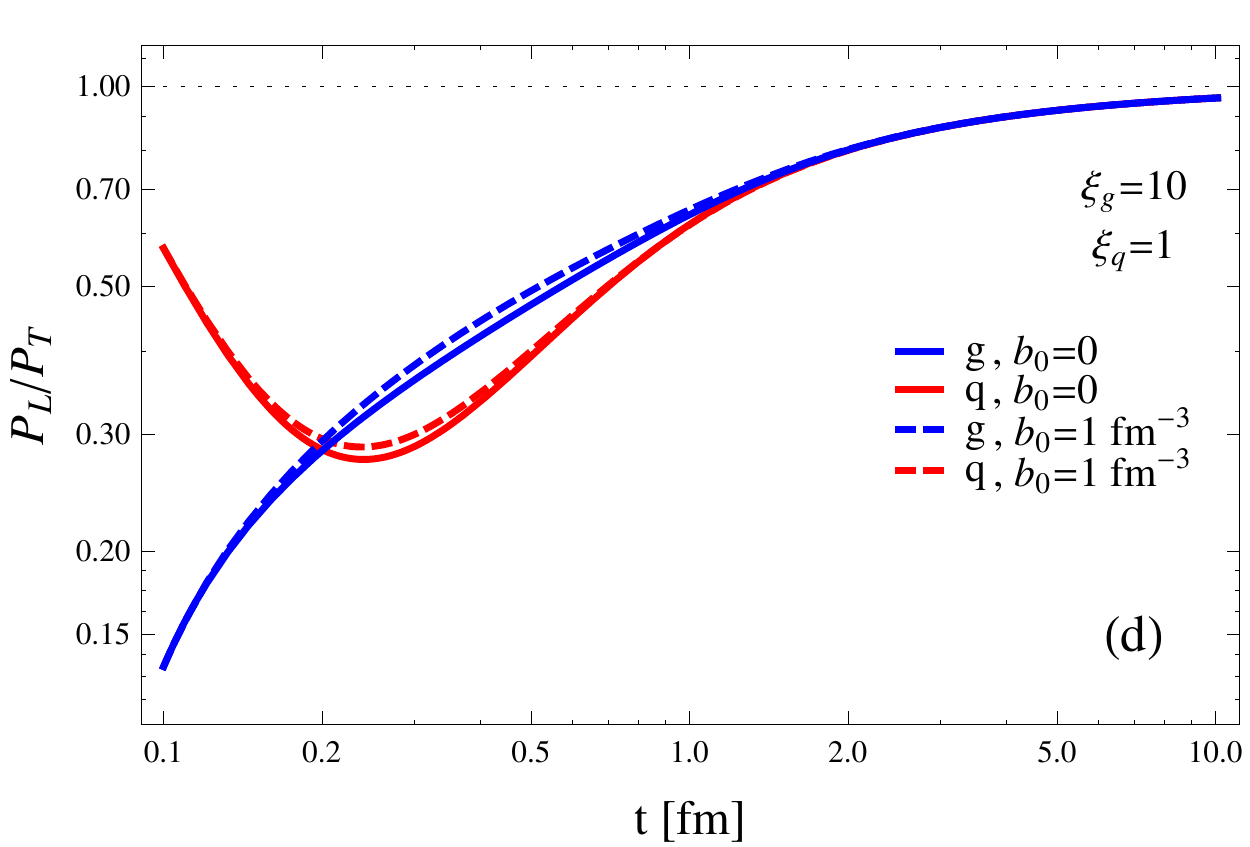}}
\end{center}
\caption{(Color online) The same as Fig.~\ref{fig:oo} but for the zero (solid lines) and finite (dashed lines) initial baryon density $b_0=1\hbox{ fm}^{-3}$, oblate-oblate configuration with $\xi_g =10$ and $\xi_q=1$.
}
\label{fig:ooFBD}
\end{figure}

%%%%%%%%%%%%%%%%%%%%%%%%%%%%%%%%%%%%%%%%%%%
\section{Results}
\label{sect:res}
%%%%%%%%%%%%%%%%%%%%%%%%%%%%%%%%%%%%%%%%%%%%

The cornerstone of our approach is the Landau matching condition which yields the equation for the effective temperature. In the case of zero baryon density, the Landau matching condition (\ref{LM2}) is reduced to the form
\begin{eqnarray}
T^4(\tau) &=&
\frac{D(\tau,\tau_0)}{2 (2 g_q + g_g)} \Bigg[
2 g_{q}  \Lambda_{q}^4\ \mathcal{H}\Bigg(\frac{\tau_0}{\tau}\frac{1}{\sqrt{1+\xi_{q}}}\Bigg) 
+ g_{g} \Lambda_{g}^4\ \mathcal{H}\Bigg(\frac{\tau_0}{\tau}\frac{1}{\sqrt{1+\xi_{g}}}\Bigg)\Bigg]
\nonumber\\ 
&& \hspace{2cm} +  
\int_{\tau_0}^{\tau} 
\frac{d\tau'}{2\tau_{\rm eq}}
D(\tau,\tau') 
T^{\prime \, 4} \,
\mathcal{H}\Bigg(\frac{\tau'}{\tau}\Bigg).
\label{LM0} 
\end{eqnarray}
We solve this equation numerically and obtain the function $T(\tau)$. Knowing the effective temperature, we calculate further observables like the energy density and the two pressures. In order to directly compare the results of the kinetic theory and anisotropic hydrodynamics, we construct different ratios of these quantities.

Since the scheme of anisotropic hydrodynamics assumes that the transverse momentum scales are the same for all the components of the mixture, in the kinetic calculations we assume that the initial scales $\Lambda_q$ and $\Lambda_g$ are the same. We note that a particular value of $\Lambda_q$ is irrelevant for our study, as all physical observables  scale with the appropriate power of $\Lambda_q$ and this parameter cancels in the ratios. The initial time for solving the kinetic and hydrodynamic equations is $\tau_0 =$ 0.1 fm/c and the relaxation time is kept constant $\tau_{\rm eq} =$ 0.25 fm/c. 

In the four panels of Figs.~\ref{fig:oo}--\ref{fig:pp} we show the results obtained in the frameworks of the kinetic theory and anisotropic hydrodynamics for the ratio of the transverse pressure to the energy density, $P_T/\varepsilon$, and for the ratio of the two pressures, $P_L/P_T$. The two upper panels show the ratios $P_T/\varepsilon$ and $P_L/P_T$ where $\varepsilon$, $P_T$, and $P_L$ describe the whole system ($\varepsilon$, $P_T$, and $P_L$ are sums of the quark and gluon contributions). The two lower panels show the ratios $P_T/\varepsilon$ and $P_L/P_T$ for the quark and gluon contributions separately. The solid lines show always the results of the kinetic theory, while the dashed lines represent the anisotropic hydrodynamics results for the quark-gluon mixture.

In Fig.~\ref{fig:oo} we present an example of our results for the case where the two initial anisotropy parameters are positive, hence, the two initial momentum distributions are oblate. In this case we find a rather good agreement between the kinetic calculation and the anisotropic hydrodynamics result. One can notice that the agreement is better for the global quantities (which characterize the sums of quarks and gluons, shown in the two upper panels of Fig.~\ref{fig:oo}) than for the individual contributions (shown in the two lower panels of Fig.~\ref{fig:oo}).

In Fig.~\ref{fig:op} we present an example of our calculations for the case where one anisotropy parameter is positive while the second one is negative. This corresponds to a mixed oblate-prolate initial configuration in the momentum space. Compared to the oblate-oblate case presented in Fig.~\ref{fig:oo} we find now much worse agreement. The solutions of the kinetic equation and anisotropic hydrodynamics differ in a qualitative way. In the framework of anisotropic hydrodynamics for mixtures, it has been found that the initial oblate-prolate and prolate-prolate configurations lead to the exponential decay of initial anisotropies. This is indicated by the dashed lines in Fig.~\ref{fig:op} which tend fast to the equlibrium values (1/3 for $P_T/\varepsilon$ and 1 for $P_L/P_T$). This type of behavior is not supported by the kinetic calculations. The critical behavior and exponential decays of the anisotropy parameters in anisotropic hydrodynamics were connected with the highly non-linear structure of the hydrodynamic equations. Such critical behavior is not expected on the grounds of the kinetic theory where all quantities depend very smoothly on the evolution time.

In Fig.~\ref{fig:pp} we show our results for the initial prolate-prolate configurations. Similarly to the prolate-oblate case, there are qualitative differences in the time behavior of the system. Again, the anisotropy parameters in the hydrodynamic approach decay exponentially to zero, while the kinetic theory  results show the power law behavior.

The presented results show limitation in the application of the anisotropic hydrodynamics for the mixture of fluids with initial oblate-prolate or prolate-prolate configurations. Clearly, the approximations used in the anisotropic hydrodynamics are too restrictive. We recall that the approach introduced in Ref.~\cite{Florkowski:2012as} was based on the zeroth and first moments of the kinetic equations and, consequently, the number of independent parameters in the ansatz for the anisotropic distribution functions was restricted. In particular, the transverse momentum scales in the quark and gluon distributions were set to be equal for the whole evolution time of the system. Our present calculation indicates that such an assumption must be relaxed and, probably, higher moments of the kinetic equation should be used to derive extra equations that allow for the determination of the independent evolution of these two scales.

Finally, in Fig.~\ref{fig:ooFBD} we compare the results obtained for vanishing and finite initial baryon density, both obtained from the kinetic equation. The zero-baryon density results are shown as the solid lines and the finite density results are represented by the dashed lines. The initial baryon density is \mbox{$b_0=1\hbox{ fm}^{-3}$}. Other parameters are the same as those used in Fig.~\ref{fig:oo}. We find that the effects of finite baryon density are negligible even for relatively high initial baryon density as used in the presented calculation.

%%%%%%%%%%%%%%%%%%%%%%%%%%%%%%%%%%%%%%%%%%%
\section{Conclusions}
\label{sect:concl}
%%%%%%%%%%%%%%%%%%%%%%%%%%%%%%%%%%%%%%%%%%%%

In this paper we have derived and analyzed a system of coupled kinetic equations for quarks, antiquarks and gluons. The results of the kinetic equations have been compared to the predictions of the anisotropic hydrodynamics describing a mixture of anisotropic fluids. We have found that the solutions of the kinetic equations can be well reproduced by anisotropic hydrodynamics in the case where the initial distribution are oblate for both quarks and gluons. Especially, the good agreement is found if one looks at the global characteristics of the system such as the total energy density or total longitudinal and transverse pressures. On the other hand, the solutions of the kinetic equations have a different qualitative behavior from those obtained in anisotropic hydrodynamics if the initial configurations are oblate-prolate or prolate-prolate. In the latter case, the solutions of anisotropic hydrodynamics exhibit exponential behavior which is absent in the solutions of the kinetic equations. Our study suggets, that the assumption of equal transverse-momentum scales for quarks and gluons used in the derivation of the hydrodynamic equations in \cite{Florkowski:2012as} should be relaxed and an additional equation should be added to the scheme from the second moment of the kinetic equation.

\medskip
We thank Mike Strickland and Radek Ryblewski for useful discussions and comments. This work was supported in part by the Polish National Science Center with Decision No. DEC-2012/06/A/ST2/00390.

\end{document}